
\tolerance=1000000
\def\ninit{\hoffset=.50 truecm
            \voffset=0. truecm
            \hsize=14.5 truecm
            \vsize=23.5 truecm
            \baselineskip=19pt
            \lineskip=0pt
            \lineskiplimit=0pt}

\def\pag{\pageno=2\footline={\hss\tenrm\folio\hss}}
\def\oneskip{\vskip\baselineskip}

\def\frac#1/#2{\leavevmode\kern.1em
\raise.5ex\hbox{\the\scriptfont0 #1}\kern-.1em
/\kern-.15em\lower.25ex\hbox{\the\scriptfont0 #2}}

\def\lsim{\, \lower2truept\hbox{${<
\atop\hbox{\raise4truept\hbox{$\sim$}}}$}\,}
\def\gsim{\, \lower2truept\hbox{${>
\atop\hbox{\raise4truept\hbox{$\sim$}}}$}\,}

\ninit
\nopagenumbers
\null
\centerline{\bf THE EXTRAGALACTIC INFRARED BACKGROUND }

\oneskip
\oneskip
\parindent=25pt
\parskip 0pt
\centerline{{G. De Zotti$\ ^1$\/}, {A. Franceschini$\ ^1$\/},
{P. Mazzei$\ ^1$\/}, {L. Toffolatti$\ ^1$\/}, and {L. Danese$\ ^2$\/}}

\oneskip
\noindent{$^1$\it Osservatorio Astronomico,
Vicolo dell'Osservatorio 5, I-35122 Padova, Italy}

\oneskip
\noindent{$^2$\it Dipartimento di Astronomia, Universit\`a di Padova,
Vicolo dell'Osservatorio 5, I-35122 Padova, Italy}

\oneskip
\oneskip
\oneskip
\oneskip

\parskip=0pt
Address for correspondence: Prof. Gianfranco De Zotti

\parindent=5.3truecm
Osservatorio Astronomico di Padova

Vicolo dell'Osservatorio 5

I-35122 Padova -- Italy

\oneskip
\oneskip

\noindent
Send offprint requests to: G. De Zotti

\vfill\eject

\pag
\centerline {\bf Abstract}
\oneskip
\parindent=1truecm
\noindent
Current limits on the intensity of the extragalactic infrared background
are consistent with the expected contribution from evolving galaxies.
Depending on the behaviour of the star formation rate and of the initial mass
function, we can expect that dust extinction during early evolutionary phases
ranges from moderate to strong. An example of the latter case may be the
ultraluminous galaxy IRAS F$10214 + 4724$. The remarkable lack of high
redshift galaxies in faint optically selected samples may be indirect
evidence that strong extinction is common during early phases. Testable
implications of different scenarios are discussed; ISO can play a key role
in this context. Estimates of possible contributions of galaxies to the
background under different assumptions are presented. The COBE/FIRAS
limits on deviations from a blackbody spectrum at sub-mm wavelengths
already set important constraints on the evolution of the far-IR emission
of galaxies and on the density of obscured (``Type 2'') AGNs. A major
progress in the field is expected at the completion of the analysis of
COBE/DIRBE data.

\vfill\eject

\noindent
{\bf Observational constraints }

\medskip\noindent
Observations of the extragalactic IR background are severely hampered
by the presence of bright foregrounds. Even when measurements are made
from space with cryogenically cooled instruments, and in favourable directions
(far from the galactic plane and from the ecliptic plane)
the extragalactic flux is overwhelmed by zodiacal light and starlight
at near-IR wavelengths, by interplanetary dust emission in the mid-IR, and
by interstellar dust emission in the far-IR/sub-mm.

The available observational information on extragalactic background radiations
over the full range from radio to $\gamma$-rays is summarized in Fig.~1.
At sub-mm wavelengths, tight upper limits come from the
absence of detectable deviations from a blackbody spectrum in the COBE/FIRAS
measurements (Mather {\it et al.} 1994). The derivation of accurate limits to
the extragalactic background is complicated by the fact that its spectrum in
this region is similar to that of the emission from our own Galaxy, so that
there is some ambiguity in the subtraction of the latter. The dot-dashed line
shows the estimate of Wright {\it et al.} (1994), derived assuming that the
Galactic emission obeys a simple cosecant law.

At higher frequencies, we have plotted as upper limits the
total observed sky brightnesses at different frequencies
measured by COBE/DIRBE in a dark direction (Hauser {\it et al.} 1991),
since it is well known that most of the observed flux is due to foreground
emission. These limits will obviously become much tighter when the analysis
of these data and the subtraction of foregrounds will be completed.

Also shown are the results of the rocket experiment by Lange
{\it et al.} (1990) at 100, 135 and $275\,\mu$m, for which a subtraction
of the foreground emission has been carried out and an indication of a truly
extragalactic signal at the intermediate wavelength has been reported.

The estimated contributions of zodiacal light, interplanetary dust emission
and light from unresolved Galactic stars have also been subtracted from the
total near-IR sky brightness observed with the rocket-borne experiments
of Matsumoto {\it et al.} (1988) and Noda {\it et al.} (1992).
Using an improved model of the Galaxy, Franceschini {\it et al.} (1991a)
concluded that the starlight contribution might have been
underestimated by Noda {\it et al.} (1992) and that
only upper limits could be set on any residual, possibly extragalactic,
component; the derived limits are plotted in Fig. 1.
Matsumoto {\it et al.} (1988)
reported a possible measurement of the extragalactic component at $2.28\,\mu$m
[not confirmed by Noda {\it et al.} (1992)] and at $3.8\,\mu$m
(plotted in Fig. 1).

As pointed out by Stecker {\it et al.} (1992), observations of very high-energy
(TeV) $\gamma$-rays in the spectra of blazars provide interesting constraints
on the intensity of the extragalactic IR background. The point is that
a flux of extragalactic $\gamma$-rays is attenuated by interactions with
ambient photons leading to the production of an electron-positron pair,
occurring primarily when the product of the two photon energies is
$\approx 0.5\,\hbox{MeV}^2$; for TeV photons, the attenuation is thus primarily
due to IR photons. The method has been applied by Stecker \& de Jager (1993),
de Jager {\it et al.} (1994) and Dwek \& Slavin (1994)
by exploiting the detection
of the BL Lac object Mrk~421 at TeV energies using ground based Cherenkov
detectors. These data show
some evidence of a spectral cutoff at a few TeV; the corresponding value
of the optical depth to pair production may comprise contributions intrinsic
to the source and therefore translates in an upper limit on the energy
density of the IR background. Shown in Fig. 1 (short-dashed curve) are
limits derived by Dwek \& Slavin (1994).

A lower limit to the background flux at $2.2\,\mu$m (also shown in Fig. 1)
comes from an integration of the deep K-band counts of galaxies
(Gardner {\it et al.}
1994). Since these counts are already relatively flat at the faintest
magnitudes, the additional contribution from galaxies below the detection
limit is probably not very large.

Altogether, simple estimates consistent with these limits (dashed curve in
Fig. 1, with the peak at $\sim 100\,\mu$m due to dust emission and the peak
at $\sim 1\,\mu$m due to starlight) suggest a global contribution of galaxies
to the extragalactic energy density in the range $1\,$mm--$1\,\mu$m of
$$\epsilon_{{\rm IR}}(z=0)\approx
1.2\times 10^{-14}\,\hbox{erg}\,\hbox{cm}^{-3},$$
within a factor of a few.
This is much higher than the energy density of the ``astrophysical''
background (as opposed to the ``cosmological''
microwave background which comprises an energy density about 30 times
larger: $\epsilon_{{\rm MWB}}(z=0)\approx 4.2\times 10^{-13}\,\hbox{erg}\,
\hbox{cm}^{-3}$) in any other waveband (see Fig. 1). For comparison, the
global energy density of the X-ray background is $\epsilon_{{\rm XRB}}(z=0)
\approx 8\times 10^{-17}\,\hbox{erg}\,\hbox{cm}^{-3}$. This already indicates
that nuclear activity, which yields a roughly flat
spectral energy distribution over many orders
of magnitude, is unlikely to contribute much (but see below).

\bigskip\medskip\noindent
{\bf Contribution of galaxies to the IR background: kinematic models }

\medskip\noindent
As mentioned above, the deep K-band counts (Gardner {\it et al.} 1993)
may already
have directly resolved most of the extragalactic near IR background.
On the other hand, the deepest far-IR survey available so far (Hacking 1987;
Hacking \& Houck 1987) with a flux limit of 50 mJy at $60\,\mu$m, extends only
out to a median redshift $z \simeq 0.08$ (Ashby {\it et al.} 1994).
Nevertheless,
the IRAS $60\,\mu$m counts have provided some evidence of a substantial
cosmological evolution of galaxies in the far-IR, consistent with that
inferred from VLA surveys at sub-mJy levels (Windhorst {\it et al.} 1993 and
references therein), given the tight correlation between radio and far-IR
emissions of disk and irregular galaxies (Hacking {\it et al.} 1987;
Danese {\it et al.} 1987; Lonsdale \& Hacking 1989; Rowan-Robinson
{\it et al.} 1993).

Simple estimates of the integrated far-IR emission of galaxies have been
produced by several authors adopting empirical evolution models
for the local luminosity function, consistent with the IRAS counts
(Franceschini {\it et al.} 1991b; Hacking \& Soifer 1991; Beichman \& Helou
1991; Oliver {\it et al.} 1992). More physical models have been considered
by Wang (1991a,b)
who has taken into account the evolution of the dust mass in galaxies
(assumed to be proportional to the mass of interstellar gas and to its
metallicity) and of the dust emissivity (the total power emitted by
each grain was taken to be proportional to the star formation rate).
Treyer \& Silk (1993) have exploited the kinematic evolution models by Cole
{\it et al.} (1992), matching optical and near-IR counts of galaxies and
redshift distributions. One model includes a new population of low luminosity
blue galaxies showing up at $z \gsim 0.7$ and gradually vanishing
at lower redshifts, because of rapid fading, merging or self-disruption.
In a second model, the evolution of dwarf blue galaxies follows the
evolution of the number density of dark matter halos in the cold dark matter
scenario, computed using the Press \& Schechter (1974) formalism.
The models have been extended throughout the electromagnetic spectrum,
from radio to X-rays, using observed correlations to relate the blue luminosity
to emissions in other wavebands. The far-IR emission of local galaxies
is assumed to be non-evolving, while that of dwarf blue galaxies is calculated
as $L_{\rm fir} = [\exp(\tau) - 1]L_{\rm B}$; results are presented
for a constant optical depth $\tau = 1$.

A thorough treatment should take into account:
the evolution of stellar populations in galaxies of different
morphological types;
the evolution of dust properties, abundance and distribution and the
corresponding evolution of extinction of starlight and re-emission
in the far-IR; the observed properties (spectral energy distributions and
luminosity functions) of galaxies
of different morphological types in the local universe.

Important constraints are provided by the deep counts in the optical
and near-IR bands, by the IRAS $60\,\mu$m counts and (thanks to the
tight far-IR/radio correlation for galaxies) by the sub-mJy VLA counts
(probably dominated by galaxies with active star formation, cf. Benn et
al 1993), as
well as by related statistics (luminosity, redshift and color distributions).

\bigskip\noindent
{\it A simple self-consistent model}

\medskip\noindent
Mazzei {\it et al.} (1992, 1994) were the first to attempt an extension up to
far-IR/sub-mm wavelengths of population synthesis models for the
chemical and photometric evolution of galaxies. Two extreme cases were
analyzed. On one side, disk galaxies,
characterized by dissipational
collapse, with slow gas depletion, i.e. the star formation rate
(SFR) never much higher than today (Sandage 1986). At the other side,
spheroidal galaxies thought
to have used up most of their gas to form stars in a time short compared
with the collapse time, i.e. with a spectacularly large
initial SFR.

For each case, chemical and photometric evolution models were constructed
in the usual way, adopting a Schmidt law for the evolution of the star
formation rate and the standard Salpeter or Scalo Initial Mass Functions
(IMFs). Simple assumptions were adopted for the dust component, i.e.:

\item{i)} the dust to gas ratio is proportional to some power of
the metallicity, as proposed by Guiderdoni \& Rocca-Volmerange (1987);

\item{ii)} stars and dust are well mixed;

\item{iii)} the ``standard'' grain model (Mathis {\it et al.} 1977; Draine
\& Lee 1984),
including a power law grain size distribution, holds at any time.

\medskip\noindent
The local optical depth is obtained
from the condition that  the amount of absorbed starlight
equals the far-IR emission; its evolution follows directly from that
of the gas fraction and of the metallicity, computed using the standard
equations for chemical evolution for given SFR and IMF.

The models allow for two dust components: cold dust, heated by the general
interstellar radiation field, and warm dust associated to starforming regions.
They include PAH molecules (Puget \& L\'eger 1989)
and emission from circumstellar dust shells,
primarily associated to stars in the final stage of evolution along the
asymptotic giant branch.

The standard view that the metallicity and the star formation rate
in galactic disks do not vary much after a few Gyr from the formation of the
disk imply that both the bolometric luminosity of these galaxies and
the re-emission of their interstellar dust vary
only slowly throughout most of the galaxy lifetime (Fig. 2).
Such weak evolution cannot account for the substantial cosmological
evolution indicated by the deep IRAS counts and by the sub-mJy
radio counts (linked to far-IR counts by the tight
far-IR/radio correlation exhibited by galaxies).

On the contrary, dramatic far-IR evolution is expected for
early-type galaxies due to the fast (exponential with a timescale of a few
Gyr) decrease of the SFR with increasing galactic age. On one hand,
the bolometric luminosity increases by a substantial factor with
decreasing galactic age, $T$: models generally
indicate a factor $\simeq 10$ increase of $L_{\rm bol}$ from $T = 15\,$Gyr to
$T = 2\,$Gyr). On the other hand,
the far-IR to optical luminosity ratio increases from local
values
$\lsim 10^{-2}$ (Mazzei \& De Zotti 1994a)
to $\simeq 1$ or even $\gg 1$ at early times.

Under the assumptions listed above, the key parameter is the gas
consumption rate: in the case of a fast conversion of gas into stars,
the far-IR emission is never dominant; but if the gas depletion is slower
the galaxy may experience a prolonged opaque phase, with most of the
luminosity emitted in the far-IR (Fig. 3). Examples of evolution of
the spectral energy distribution of spheroidal galaxies predicted by
different models are shown in Fig. 4.

\bigskip\noindent
{\it An optically thick early evolutionary phase of spheroidal galaxies?}

\medskip\noindent
It has been stressed by many authors (e.g. van den Berg 1990, 1992; Wang
1991 a,b,c; Kormendy and Sanders 1992) that, due to the fast metal
enrichment of primordial galaxies, substantial
dust extinction and far-IR emission are
expected during early phases of galaxy evolution.

As illustrated by Fig. 3,
if the dust to gas ratio is roughly proportional to the metallicity,
the effective dust opacity may have very different histories,
depending on the behaviour of the SFR, of the IMF, but possibly
also on external effects such as infall of intergalactic material
or merging of gas rich companions. In particular,
if the gas depletion is relatively slow, the galaxy may experience an
extremely optically thick phase. An example of the
latter situation may be the ultraluminous galaxy IRAS F10214$+$4724
(Rowan-Robinson {\it et al.} 1991; Lawrence {\it et al.} 1993;
Mazzei \& De Zotti 1994b;
see Fig. 5).

Under some circumstances (see Fig. 3), the strongly optically thick
phase may last several Gyr. It is presently unknown how frequently
such situation may occur. However, some recent results provide
indirect indications that strong extinction may indeed be common
during early evolutionary phases (Franceschini {\it et al.} 1994).

The standard population synthesis models for galaxy evolution
(Guiderdoni \& Rocca-Volmerange 1990; Yoshii \& Takahara 1988; Charlot \&
Bruzual 1991) predict that galaxies, and particularly early-type
galaxies, should have been brighter in the past
as a consequence of a more active star formation associated to a
larger gas fraction. Thus, the excess surface density of galaxies, in
comparison to expectations in the absence of evolution, was attributed
to the brightening of high-$z$ galaxies, which increases the sampled volume.

However, spectroscopic work has shown that the excess faint counts are rather
due to dwarf galaxies at relatively low redshifts ($z \sim 0.3$--0.7) and that
there is a remarkable lack of high redshift galaxies in optically
selected samples down to $B= 24$--25 (Broadhurst {\it et al.} 1988; Colless
{\it et al.} 1990, 1993; Cowie {\it et al.} 1991). In fact,
none of the 100 (out of a total of 104) galaxies with measured redshifts
in the complete sample with $B < 22.5$ of Colless {\it et al.} (1993)
has $z > 0.7$;
the highest redshift of the complete sample by Cowie {\it et al.} (1991) with
$B < 24$ is $z=0.73$. Colless {\it et al.} (1993) conclude that their data
allow no more than 1 mag of luminosity evolution of $L_\star$ galaxies by
$z=1$.

An attractive explanation of these results is
that present day luminous galaxies formed late as the
result of merging, perhaps of the same faint blue galaxies which appear
to be much more numerous at intermediate redshifts than they are locally.
This hypothesis, however, faces some serious problems:

i) the thinness of disks implies that only a few percent of the baryonic
mass can have been accreted in the last several Gyr (Ostriker 1990; Toth \&
Ostriker 1992; Quinn {\it et al.} 1993);

ii) colors and the existence of a well defined ``fundamental
plane'' for early type galaxies provide strong indications that they
are old ($z_F > 2$) and essentially coeval, later additions probably
corresponding to no more than 10\% of the present luminosity (Renzini 1993);

ii) faint galaxies in the relevant magnitude range ($22 < B < 26$) appear
to be
less spatially clustered than local galaxies, contrary to expectations if
they are in the process of merging (Efstathiou {\it et al.} 1991; Pritchet \&
Infante 1992; Roche {\it et al.} 1993);

iii) the merging timescale needed to explain the faint galaxy
counts appears to be uncomfortably higher than current estimates based on
local galaxy samples (Carlberg 1992);

iv) the total luminosity in local ellipticals
is lower than expected if they are the merger products of the ``excess''
faint blue galaxies with $B < 24$ (Dalcanton 1993).

On the other hand, evidences of substantial
amounts of dust at high $z$ (up to $z = 4.69$) were provided by
detection of QSOs at mm wavelengths (Andreani {\it et al.} 1993;
McMahon {\it et al.} 1994; Isaak {\it et al.} 1994). Also,
enormous dust masses are indicated by sub-mm detections
(Hughes {\it et al.} 1994) of the
high $z$ radio galaxies 4C41.17 ($z = 3.8$) and 53W002 ($z=2.39$).

A consistent picture obtains assuming that, during the phases of intense
star formation, most of the optical radiation was absorbed by dust and
reradiated in the far-IR (Franceschini {\it et al.} 1994). In this case,
less than 1\% of galaxies in the range $21 < B < 22.5$ and less than
10\% of those with $B < 24$ are expected at $z > 0.7$ (in the absence of
dust extinction, the standard evolutionary models predict that 40\% of
galaxies with $21 < B < 22.5$ and $\sim 50\%$ of those with $B < 24$
should have $z > 0.7$). Moreover,
the model can account for the deep $60\,\mu$m IRAS counts and,
exploiting the far-IR/radio correlation for galaxies, can explain most
of the observed sub-mJy flattening of radio counts over a couple of decades
in flux.
Also, dust extinction may explain the failure to detect Ly$\alpha$
emission in searches for primeval galaxies (De Propris {\it et al.} 1993;
Djorgovski \& Thompson 1992).

This scenario entails a number of testable predictions:

i) the redshift distribution of galaxies in the deep ($S_{60\mu m} >
50\,$mJy) IRAS sample (Hacking \& Houck 1987; Hacking {\it et al.} 1987;
Ashby {\it et al.} 1994)
should have a tail at significant $z$: we expect $\sim 30$--40 galaxies
at $z > 0.1$ and about 5-10 at $z > 0.3$; in the absence of
evolution, only about 18 should be at $z > 0.1$ and essentially none at
$z > 0.3$;

ii) a high redshift peak should quickly develop in the redshift distribution
at fainter flux limits, easily reachable with ISO;

iii) a substantial fraction of high-$z$ galaxies should also be present
among sub-mJy radio sources;

iv) large ellipticals should virtually disappear at $B \simeq 22$--24, an
effect that could be tested by morphological studies with HST.

\medskip\noindent
Estimates of possible contributions of galaxies to the background,
under different assumptions (no evolution, moderately opaque case,
opaque case), are shown in Fig. 6. The expected contribution to the near-IR
background is relatively insensitive to the effect of dust extinction.
The measurements of the expected backgrounds are well within the DIRBE
sensitivity limits. The great difficulty, however, is the subtraction
of foreground emissions.

\bigskip\medskip\noindent
{\bf Pre-galactic star formation}

\medskip\noindent
The possibility that
early structures, at $z \sim 5$--100, could have led to copious star
formation, producing both an intense background and dust capable of
reprocessing it, has been extensively discussed by Bond {\it et al.} (1991).
In this case, essentially all the energy produced by nuclear reactions
comes out at far-IR/sub-mm wavelengths. The peak wavelength depends
on the redshift and temperature distributions of the dust but, for
a relatively broad range of parameter values, occurs at $\lambda
\sim 600\,\mu$m. As shown by Wright {\it et al.} (1994), only one of the many
models worked out by Bond {\it et al.} (1991) is compatible with the COBE/FIRAS
data.

\bigskip\medskip\noindent
{\bf Constraints on dust enshrouded AGN populations }

\medskip\noindent
As already mentioned, canonical AGNs probably make a minor contribution to the
IR background. In fact, quasars were very rarely detected in the IRAS survey.
The list of previously unknown quasars discovered by IRAS, compiled by
Clowes {\it et al.} (1991), contains only eleven objects,
a number comparable to that
of previously cataloged quasars included in the IRAS Point Source Catalog
(cf. Soifer {\it et al.} 1987). The question then arose whether they constitute
a new population of infrared loud quasars, perhaps corresponding to
an early phase of the evolution, when the active nucleus was enshrouded
by a dense cloud of dust and gas which is gradually consumed and/or swept
away as the luminosity of the nucleus increases (Sanders {\it et al.} 1988; Low
{\it et al.} 1988). The work of Low {\it et al.} (1989),
Sanders {\it et al.} (1989) and
Clowes {\it et al.} (1991), however, suggests that IRAS selected quasars
are not very special compared with other quasars; they simply correspond to
the reddest tail of the general population, as expected as a consequence of the
far-IR selection, and are mainly demostrating the incompleteness of the
optical bright quasar surveys.

On the other hand, AGNs
have been so far identified through their optical properties.  It is
still possible that there exists a significant population of heavily obscured
AGNs which are optically very faint, but bright in the far-IR.
It has been argued that dust enshrouded AGNs can
contribute significantly to power the highest luminosity IRAS galaxies
(e.g. Sanders {\it et al.} 1987). Two very well studied examples of very high
infrared luminosity galaxies are Arp 220 (IC 4553; Soifer {\it et al.} 1984)
and NCG 6240 (Wright {\it et al.} 1984);
both show strong evidence for an active nucleus.
The extreme far-IR galaxies IRAS F10214$+$4724 (Rowan-Robinson {\it et al.}
1991; Lawrence {\it et al.} 1993; Elston {\it et al.} 1994),
IRAS F15307$+$3252 (Cutri et
al 1994), and IRAS 09104$+$4109 (Kleinman \& Keel 1987; Kleinmann et
al 1988; Hines \& Wills 1993) all clearly harbor an AGN (although its
relative contribution to the bolometric luminosity is still unclear).

In a different context, extreme absorption has been postulated by Setti \&
Woltjer (1989), Morisawa {\it et al.} (1990), and Grindlay \& Luke (1990) to
account for the spectrum of the X-ray background above $\simeq 3$ keV, in the
framework of unified models for AGNs. According to these
models, the central powerhouse is surrounded, at a distance of
several parsecs, by obscuring matter, probably having toroidal geometry.
Depending on whether our view is down the axis or is occulted by the torus,
we see a broad line AGN (type 1 Seyfert or QSO) or a narrow line one
(type 2 Seyfert or hypothetical type 2 QSO), respectively.
For reasonable values of tori masses and sizes, absorbing columns up to
$10^{24}$--$10^{25}\,\hbox{cm}^{-2}$ can be expected, implying that the
nuclear X-ray emission can be
absorbed up to $\sim 20$--30 keV, depending on the torus column density
and geometry. The density of strongly absorbed sources required
to account for the X-ray background above 3 keV is a few times larger than
that of type 1 AGNs and their cosmological evolution must be similarly
strong (Comastri {\it et al.} 1994; Madau {\it et al.} 1993, 1994).

If such highly obscured AGNs emit most of their power in the far-IR, their
contribution to the far-IR background may be substantial, as illustrated by the
following simple calculation.
The local energy density produced by UV-excess AGNs can be
estimated from B counts as:
$$\epsilon_{{\rm AGN}}\simeq {4\pi\over c}\kappa\cdot I_B\simeq {4\pi \over c}
\kappa\nu_B\int_{S_{min}}^{S_{max}} S\cdot N_B(S)dS\  .$$
\noindent
Assuming a ratio of bolometric to B-band flux $\kappa = 30$ (Padovani 1989)
and an effective redshift $z_{{\rm AGN}} = 1.5$
we find $\epsilon_{{\rm AGN}}(z=0)\approx
1.5\times 10^{-15}\,\hbox{erg}\,\hbox{cm}^{-3}$, not far from the
estimated contribution of galaxies to the background at $\lambda > 10\,\mu$m
($\epsilon_{{\rm gal, FIR}}(z=0)\approx
4\times 10^{-15}\,\hbox{erg}\,\hbox{cm}^{-3}$).

Since the estimated contribution from galaxies is already close to
the COBE/FIRAS limits at sub-mm wavelengths, these limits set
significant constraints on
a population of dust-enshrouded AGNs, with a bolometric emission
comparable to that of canonical AGNs. If they are to account for the hard
X-ray background, they would yield a contribution to the far-IR
background potentially detectable by COBE/DIRBE.

Note that the quoted radiation energy density due to optically selected
AGNs corresponds to a
quite low mass density of collapsed nuclei: $(H_0/50)^2 \Omega_{{\rm AGN}}
\simeq 5.5\times 10^{-6}(\kappa/30)(\eta/0.1)^{-1}$, $\Omega$ being the
mass density in units of the critical density and $\eta$ the mass to energy
conversion efficiency (Padovani {\it et al.} 1990). It follows that the
available data on diffuse backgrounds (cf. e.g. Fig. 1)
constrain the mass density that can be contributed by black holes built
up by accretion with high radiation efficiency. In particular, the limits
on the extragalactic far-IR background (taking the Oliver {\it et al.} (1992)
100 $\mu$m
upper limit as a reference value) imply that the mass density
of dust enshrouded AGNs has to be $(H_0/50)^2 \Omega_{{\rm AGN}}
\lsim 3.3\times 10^{-5}(\eta/0.1)^{-1}(\kappa/30)$.

\bigskip\noindent
{\bf Conclusions}

\medskip\noindent
There are indications that a large fraction of starlight is absorbed
by dust and re-radiated at far-IR wavelengths.
If so, surveys at longer wavelengths, where the effect of dust
extinction is smaller, should contain higher fractions of high-$z$ galaxies.
A hint in this direction might be seen in the I-band sample of Lilly (1993),
where 6 out of 25 confirmed galaxies are at $z > 0.8$.

At mid-IR wavelengths three effects concur in easing the detection
of high-z galaxies: the decreasing effect of dust extinction; the
positive K-correction (the stellar SED peaks at $\lambda \simeq 1\,\mu$m);
the positive luminosity evolution. The deep survey with
the most sensitive filter of ISOCAM (LW2, $5 < \lambda < 8.5\,\mu$m)
should detect $\sim 2$--3 galaxies per arcmin$^2$, i.e. 10--20 galaxies
per frame; a good fraction of these should be at $z \gsim 1$.

Our models imply that early type galaxies with intense star formation
could start to show up in $60\,\mu$m surveys at $S_{60\mu m} \lsim 100\,$mJy.
A few galaxies in the deep ($S_{60\mu m} > 50\,$mJy) sample by
Hacking {\it et al.} (1987) are expected to be early type galaxies at $z >
0.6$.

ISO should be able to cover a few deg$^2$ to $S_{60\mu m} \sim 10$ mJy;
various tens of dusty galaxies may be detected; we expect that a good
fraction of them are at significant redshifts.

Some of the galaxies detected in the deepest (sub-mJy) radio surveys
may also be at significant $z$. They are expected to be very faint in the
optical (up to $B \simeq 28$).

Optically selected
AGNs are minor contributors to the far-IR background. A substantial
contribution, however, may be expected from a yet undetected population
of highly obscured nuclei, such as that advocated to explain the
X-ray background above 3 keV.

Tight constraints on the evolution of dusty
galaxies and on the density and evolution of dust enshrouded AGN are expected
from COBE/DIRBE data, when the delicate subtraction of foreground emission will
be completed. As shown by Wright {\it et al.} (1994), COBE/FIRAS limits on the
sub-mm background have already ruled out a large variety of models postulating
intense pre-galactic star formation (Bond {\it et al.} 1991) as well as of
non-standard models, such as those invoking early bursts
of Population III stars producing
a large fraction of the cosmic helium abundance (Negroponte {\it et al.} 1981;
Wright 1981).

\vskip 1cm
\noindent
Work supported in part by CNR, ASI and by the EC program Human Capital
and Mobility.

\vfill\eject

%
%
\def\aa #1 #2{{\it Astr. Ap.,}~{\bf #1}, {#2}}
\def\aas #1 #2{{\it Astr. Ap. Suppl.,}~{\bf #1}, {#2}}
\def\araa #1 #2{{\it Ann. Rev. Astr. Astrophys.,}~{\bf #1}, {#2}}
\def\aj #1 #2{{\it A. J.,}~{\bf #1}, {#2}}
\def\alett #1 #2{{\it Ap. Lett.,}~{\bf #1}, {#2}}
\def\apj #1 #2{{\it Ap. J.,}~{\bf #1}, {#2}}
\def\apjl #1 #2{{\it  Ap. J. (Lett.),}~{\bf #1}, L#2}
\def\apjs #1 #2{{\it Ap. J. Suppl.,}~{\bf #1}, {#2}}
\def\ass #1 #2{{\it Ap. Space Sci.,}~{\bf #1}, {#2}}
\def\baas #1 #2{{\it Bull. Am. astr. Soc.,}~{\bf #1}, {#2}}

\def\ca #1 #2{{\it Comm. Ap. Sp. Phys.,}~{\bf #1}, #2}
\def\fcp #1 #2{{\it Fundam. Cosmic Phys.,}~{\bf #1}, {#2}}
\def\memsait #1 #2{{\it Memorie Soc. astr. ital.,}~{\bf #1}, {#2}}
\def\mnras #1 #2{{\it M. N. R. A. S.,}~{\bf #1}, {#2}}
\def\qjras #1 #2{{\it Q. J. R. astr. Soc.,}~{\bf #1}, {#2}}
\def\nat #1 #2{{\it Nature,}~{\bf #1}, {#2}}
\def\pasj #1 #2{{\it Publs astr. Soc. Japan,}~{\bf #1}, {#2}}
\def\pasp #1 #2{{\it Publs astr. Soc. Pacif.,}~{\bf #1}, {#2}}

\def\physl #1 #2{{\it Phys.Lett.,}~{\bf #1}, #2}
\def\physrep #1 #2{{\it Phys. Rep.,}~{\bf #1}, #2}

\def\physreva #1 #2{{\it Phys. Rev. A,}~{\bf #1}, {#2}}
\def\physrevb #1 #2{{\it Phys. Rev. B,}~{\bf #1}, {#2}}
\def\physrevd #1 #2{{\it Phys. Rev. D,}~{\bf #1}, {#2}}
\def\physrevl #1 #2{{\it Phys. Rev. Lett.,}~{\bf #1}, {#2}}
\def\pl #1 #2{{\it Phys. Lett.,}~{\bf #1}, {#2}}

\def\prsl #1 #2{{\it Proc. R. Soc. London Ser. A,}~{\bf #1}, {#2}}

\def\ptp #1 #2{{\it Prog. theor. Phys.,}~{\bf #1}, {#2}}
\def\ptps #1 #2{{\it Prog. theor. Phys. Suppl.,}~{\bf #1}, {#2}}
\def\rmp #1 #2{{\it Rev. Mod. Phys.,}~{\bf #1}, {#2}}

\def\sovastr #1 #2{{\it Sov. Astr.,}~{\bf #1}, {#2}}
\def\sovastrl #1 #2{{\it Sov.Astr. (Lett.),}~{\bf #1}, L#2}

\def\ssr #1 #2{{\it Space Sci. Rev.,}~{\bf #1}, {#2}}
\def\va #1 #2{{\it Vistas in Astronomy,}~{\bf #1}, {#2}}

\def\book #1 {{\it ``{#1}'',\ }}

\def\ref{\noindent\hangindent=20pt\hangafter=1}

\centerline{\bf References}
\bigskip

\ref
Andreani, P., F. La Franca, and S. Cristiani, \mnras 261 L35--L38 1993.

\ref
Ashby, M.L.N., P.B. Hacking, J.R. Houck, B.T. Soifer, and E.W. Weisstein,
{\it Ap. J.}, submitted, 1994.

\ref
Beichman, C.A. and G., Helou, \apj 370 L1--L4 1991.

\ref
Benn, C.R., M. Rowan-Robinson, R.G. McMahon, T.J. Broadhurst, and A. Lawrence,
\mnras 263 98--112 1993.

\ref
Bond, J.R., B.J. Carr, and C.J. Hogan, \apj 367 420--454 1991.

\ref
Broadhurst, T.J., R. Ellis, and T. Shanks, \mnras 235 827--840 1988.

\ref
Carlberg, R.G., \apj 399 L31--L35 1992.

\ref
Charlot, S., and G. Bruzual, \apj 367 126--141 1991.

\ref
Clowes, R.G., S.K. Leggett, and A. Savage, \mnras 250 597--601 1991.

\ref
Cole, S., M.A. Treyer, and J. Silk, \apj 385 9--25 1992.

\ref
Colless, M., R. Ellis, K. Taylor, and R. Hook, \mnras 244 408--420 1990.

\ref
Colless, M., R.S. Ellis, K. Taylor, and B. Peterson \mnras 261 19--38 1993.

\ref
Comastri, A., G. Setti, G. Zamorani, and G. Hasinger, {\it Astr. Ap.}, in
press, 1994.

\ref
Cowie, L.L., A. Songaila, and E.M. Hu, \nat 354 460--462 1991.

\ref
Cutri, R.M., J.P. Huchra, F.J. Low, R.L. Brown, and P.A. Vanden Bout,
\apj 424 L65--L68 1994.

\ref
Dalcanton, J.J., \apj 415 L87--L90 1993.

\ref
Danese, L., G. De Zotti, A. Franceschini, and L. Toffolatti, \apj 318 L15--L19
1987.

\ref
De Jager, O.C., F.W. Stecker, and M.H. Salamon, \nat 369 294--296 1994.

\ref
De Propris, R., C.J. Pritchet, D.A. Hartwick, and P. Hickson,
\aj 105 1243--1250 1993.

\ref
Djorgovski, G., and D.J. Thompson, in {\it The Stellar
Populations of Galaxies}, IAU Symp. 149, B. Barbuy and A. Renzini Eds.
(Kluwer, Dordrecht), 1992.

\ref
Draine, B.T., and H.M. Lee, \apj 285 89--102 1984.

\ref
Dwek, E., and J. Slavin, {\it Ap. J.}, in press, 1994.

\ref
Efstathiou, G., G. Bernstein, N. Katz, A.J. Tyson, and
P. Guhathakurta, \apj 380 L47--L50 1991.

\ref
Elston, R., P.J. McCarthy, P. Eisenhardt, M. Dickinson, H. Spinrad,
B.T. Januzzi, and P. Maloney, \aj 107 910--919 1994.

\ref
Franceschini, A., P. Mazzei and G. De~Zotti,
Proceedings of the XIth Moriond Astrophysical Meeting {\it ``The
early observable universe from diffuse backgrounds''}, B. Rocca-Volmerange,
J.M. Deharveng \& J. Tr\^an Thanh V\^an eds., Editions Fronti\`eres,
p. 249--255, 1991a.

\ref
Franceschini, A., P. Mazzei, G. De Zotti, and L. Danese,
\apj 427 140--154 1994.

\ref
Franceschini, A., L. Toffolatti, P. Mazzei, L. Danese, and
G. De Zotti, \aas 89 285--310 1991b.

\ref
Gardner, J.P., L.L. Cowie, and R.J. Wainscoat, \apj 415 L9--L12 1993.

\ref
Grindlay, J.E., and M. Luke, in {\it High Resolution X--ray Spectroscopy
of Cosmic Plasmas}, proc. IAU Coll. 115, eds P. Gorenstein
and M.V. Zombeck, Kluwer, Dordrecht, p. 276--280, 1990.

\ref
Guiderdoni, B., and B. Rocca--Volmerange, \aa 186 1--18 1987.

\ref
Guiderdoni, B., and B. Rocca--Volmerange, \aa 227 362--374 1990.

\ref
Hacking, P., Ph.D. thesis, Cornell University, 1987.

\ref
Hacking, P.B., J.J. Condon, and J.R. Houck, \apj 316 L15--L18 1987.

\ref
Hacking, P. and J.R. Houck, \apjs 63 311--330 1987.

\ref
Hacking, P.B., and B.T. Soifer, \apj 367 L49--L52 1991.

\ref
Hauser, M.G., T. Kelsall, S.H. Jr. Moseley, R.F. Silverberg,
T. Murdock, G. Toller, W. Spiesman, and J. Weiland,
in {\it Proc. AIP Conference ``After the First Three Minutes''}
eds. Holt, S.S., Bennett, C.L., Trimble, V., {\bf 222}, p.161--170 1991.

\ref
Hines, D.C., and B.J Wills, \apj 415 82--92 1993.

\ref
Hughes, D., J. Dunlop, S. Rawlings, and S. Eales, paper presented at the
European and National Astronomy Meeting, Edinburgh, 1994.

\ref
Isaak, K.G., R.G. McMahon, R.E. Ellis, and S. Withington, \mnras 269
L28--L32 1994.

\ref
Lilly, S.J. \apj 411 501--512 1993.

\ref
Kleinmann, S.G., D. Hamilton, W.C. Keel, C.G. Wynn-Williams, S.A. Eales,
E.E. Becklin, and K.D. Kuntz, \apj 328 161--169 1988.

\ref
Kleinmann, S.G., and W.C. Keel, in {\it Star Formation in Galaxies},
ed. C.J. Lonsdale, NASA Conf. Publ. 2466, 559--562, 1987.

\ref
Kormendy, J., and D.B. Sanders, \apj 390 L53--L56 1992.

\ref
Lange, A.E., P.L. Richards, S. Hayakawa, T. Matsumoto, H. Matsuo,
H. Murakami, and S. Sato, 1990, private comm. to Hauser {\it et al.}. 1991.

\ref
Lawrence, A., {\it et al.}, \mnras 260 28--44 1993.

\ref
Lonsdale, C., and P. Hacking, \apj 339 712--722 1989.

\ref
Low, F.J., R.M. Cutri, S.G. Kleinmann, and J.P. Huchra, \apj 340 L1--L4 1989.

\ref
Low, F.J., J.P. Huchra, S.G. Kleinmann, and R.M. Cutri, \apj 327 L41--L44 1988.

\ref
Madau, P., G. Ghisellini, and A.C. Fabian, \apj 410 L7--L10 1993.

\ref
Madau, P., G. Ghisellini, and A.C. Fabian, {\it M. N. R. A. S.,} in press,
1994.

\ref
Mather, J.C., {\it et al.}, \apj 420 439--444 1994.

\ref
Mathis, J.S., W. Rumpl, and K.H. Nordsieck, \apj 217 425--433 1977.

\ref
Matsumoto, T., S. Hayakawa, H. Matsuo, H. Murakami, S. Sato, A.E. Lange,
and P.L. Richards, \apj 329 567--580 1988.

\ref
Mazzei, P., and G. De Zotti, \apj 426 97--104 1994a.

\ref
Mazzei, P., and G. De Zotti, \mnras 266 L5--L9 1994b.

\ref
Mazzei, P., C. Xu, and G. De~Zotti, \aa 256 45--55 1992.

\ref
Mazzei, P., G. De Zotti, and C. Xu, \apj 422 81--91 1994.

\ref
McMahon, R.G., A. Omont, J. Bergeron, E. Kreysa, and C.G.T. Haslam,
\mnras 267 L9--L12 1994.

\ref
Morisawa, K., M. Matsuoka, F. Takahara, and L. Piro, \aa 236 299--311 1990.

\ref
Negroponte, J., M. Rowan-Robinson, and J. Silk, \apj 248 38--50 1981.

\ref
Noda, M., V.V. Christov, H. Matsuhara, T. Matsumoto, S. Matsura,
K. Noguchi, S. Sato, and H. Murakami, \apj 391 456--465 1992.

\ref
Oliver, S.J., M. Rowan-Robinson, and W. Saunders, \mnras 256 15--18P 1992.

\ref
Ostriker, J.P., in {\it Evolution of the Universe of
Galaxies}, ed R.G. Kron, p. 10--18, 1990.

\ref
Padovani, P., \aa 209 27--40 1989.

\ref
Padovani, P., R. Burg, and R.A. Edelson \apj 353 438--448 1990.

\ref
Press, W.H., and P. Schechter, \apj 187 425--438 1974.

\ref
Pritchet, C.J., and L. Infante, \apj 399 L35--L38 1992.

\ref
Puget, J.L., and A. L\'eger, \araa 27 161--195 1989.

\ref
Quinn, P.J., L. Hernquist, and D.P. Fullagar, \apj 74--93 1993.

\ref
Renzini, A., {\it Texas/Pascos '92: Relativistic Astrophysics and Particle
Cosmology}, C.W. Akerlof and M.A. Srednicki eds.,
{\it Ann. N. Y. Acad. Sci.}, {\bf 688}, 124--135, 1993.

\ref
Roche, N., T. Shanks, N. Metcalfe, and R. Fong, \mnras 263 360--368 1993.

\ref
Rowan-Robinson M. {\it et al.}., \nat 351 719--722 1991.

\ref
Rowan-Robinson, M., C.R. Benn, A. Lawrence, R.G. McMahon, and
T.J. Broadhurst, \mnras 263 123--130 1993.

\ref
Sandage, A., \aa 161 89--97 1986.

\ref
Sanders, D.B., {\it et al.}, in {\it Star Formation in Galaxies},
ed. C.J. Lonsdale, NASA Conf. Publ. 2466, 411--420, 1987.

\ref
Sanders, D.B., B.T. Soifer, J.H. Elias, G. Neugebauer, and K. Matthews,
\apj 328 L35--L38 1988.

\ref
Sanders, D.B., E.S. Phinney, G. Neugebauer, B.T. Soifer, and K. Matthews,
\apj 347 29--44 1989.

\ref
Setti, G., and L. Woltjer, \aa 224 L21--L23 1989.

\ref
Soifer, B.T., {\it et al.}, \apj 283 L1--L4 1984.

\ref
Soifer, B.T., Houck, J.R. \& Neugebauer, G., \araa 25 187--230 1987.

\ref
Stecker, F.W. and O.C. De Jager, \apj 415 L71--L74 1993.

\ref
Stecker, F.W., O.C. De Jager, and M.H. Salamon, \apj 390 L49--L52 1992.

\ref
Toth, G, and J.P. Ostriker, \apj 389 5--26 1992.

\ref
Treyer, M.A., and J. Silk, \apj 408 L1--L4 1993.

\ref
van den Berg, S., \pasp 102 503--510 1990.

\ref
van den Berg, S., \mnras  255 29--32P 1992.

\ref
Wang, B., \apj 374 456--464 1991a.

\ref
Wang, B., \apj 374 465--474 1991b.

\ref
Wang, B., \apj 383 L37--L40 1991c.

\ref
Windhorst, R.A., E.B. Fomalont, R.B. Partridge, and J.D. Lowenthal,
\apj 405, 498--517 1993.

\ref
Wright, E.L., \apj 250 1--14 1981.

\ref
Wright, E.L., {\it et al.}., \apj 420 450--456 1994.

\ref
Wright, G.S., R.D. Joseph, and W.P.S. Meikle, \nat 309 430--431 1984.

\ref
Yoshii, Y. and F. Takahara, \apj 326 1--14 1988.

\vfill\eject

\centerline{\bf  Figure Captions}

\bigskip\noindent
{\bf Figure 1.} Measurements of and constraints on the brightness per
logarithmic frequency interval of the extragalactic background from
radio to $\gamma$-ray frequencies.

\bigskip\noindent
{\bf Figure 2.} Evolution with galactic age of the spectrum of a disk
galaxy (Mazzei {\it et al.} 1992).

\bigskip\noindent
{\bf Figure 3.} Evolution of the effective optical depth, $\tau$,
of the gas metallicity, $Z$ (in solar units), and of the gas fraction, $f$,
for an early type galaxy for different choices of the SFR and of
the lower mass limit, $m_l$ (in solar masses);
see Mazzei {\it et al.} (1994) for more details. A Schmidt
parametrization has
been adopted for the SFR ($\psi(t)=\psi_0 f^n\,\hbox{M}_\odot\hbox{yr}^{-1}$).
A Salpeter form has been adopted for the IMF. If a constant fraction of metals
is locked up in dust grains, $\tau$ is proportional to $fZ$.

\bigskip\noindent
{\bf Figure 4.} Evolution with galactic age of the spectrum of an early type
galaxy (Mazzei {\it et al.} 1994), in the case of a moderate (upper panel)
or strong dust extinction during early phases. The spectrum at $T = 15$ Gyr
is compared with data for nearby ellipticals (see Mazzei {\it et al.} 1994).

\bigskip\noindent
{\bf Figure 5.} Spectral energy distribution of an early type galaxy
with strong extinction during early evolutionary phases (cf. Fig. 4),
for two values of the galactic age $T$,
compared with observational data for the galaxy IRAS F$10214+4724$
(see Mazzei and De Zotti 1994b).

\bigskip\noindent
{\bf Figure 6.} Extragalactic background light at IR to sub--mm wavelengths.
Data and upper limits are from Hauser {\it et al.} (1991), Noda {\it et al.}
(1992) as revised by Franceschini {\it et al.} (1991a),
Oliver {\it et al.} (1992) and Dwek \& Slavin (1994) (short--dashed line).
Curve $a$ correponds to no evolution, curves $b$
and $c$ to moderate or strong extinction, respectively, of early type galaxies
during early evolutionary phases; curve $d$ is the integrated starlight of
distant galaxies in the near-IR, derived from models fitting the deep
K-band counts. Curve $S$ shows the contribution of
stars in our own Galaxy, at high galactic latitudes (see Franceschini et al.
1991b). The DIRBE sensitivity as function of $\lambda$ is also shown.

\bye